\documentclass[final]{aipproc}
\layoutstyle{6x9}
\usepackage{graphicx,amsmath,wrapfig,epsfig}
\begin{document}
\title{High-energy hadron physics at J-PARC}

\author{S. Kumano}{
  address={Institute of Particle and Nuclear Studies, 
          High Energy Accelerator Research Organization (KEK) \\
           and 
           Department of Particle and Nuclear Studies,
           Graduate University for Advanced Studies \\
           1-1, Ooho, Tsukuba, Ibaraki, 305-0801, Japan}}

\begin{abstract}
The J-PARC facility is near completion and experiments
will start in 2009 on nuclear and particle physics projects.
In this article, the J-PARC facility is introduced, 
and possible projects are discussed in high-energy hadron physics
by using the primary proton beam of 30$-$50 GeV.
There are proposed experiments on charm-production and Drell-Yan
processes as well as single spin asymmetries for investigating
quark and gluon structure of the nucleon and nuclei.
Parton-energy loss could be studied in the Drell-Yan processes.
There is also a proposal on hadron-mass modifications
in a nuclear medium by using the proton beam.
In addition, possible topics include 
transition from hadron to quark degrees of freedom 
by elastic $pp$ scattering, color transparency by $(p,2p)$,
short-range correlation in nuclear force by $(p,2pN)$,
tensor structure functions for spin-one hadrons, 
fragmentation functions, and generalized parton distributions 
in the ERBL region although proposals are not written
on these projects. If proton-beam polarization is attained,
it is possible to investigate details of nucleon spin structure. 
In the last part of this article, our own studies are explained
on parton distribution functions in connection with 
the J-PARC projects.
\end{abstract}
\maketitle

\vspace{-0.8cm}
\section{Introduction}
\vspace{-0.1cm}

The J-PARC stands for the Japan Proton Accelerator Research Complex,
and it is located at Tokai in Japan \cite{j-parc}. 
It is a joint facility between JAEA (Japan Atomic Energy Agency) and
KEK (High Energy Accelerator Research Organization) for projects
in various fields of science.
The J-PARC provides most intense proton beam in the multi-GeV energy
region. Nuclear and particle physics projects use secondary beams
such as kaons, pions, and neutrinos as well as the primary 50-GeV
proton beam.
The construction is near completion and experiments will start soon.
The first nuclear and particle physics experiments are on
strangeness nuclear physics and neutrino oscillation. 

As future hadron experiments, there are many possibilities.
It is the purpose of this paper to introduce ``possible'' high-energy
hadron projects by using the primary proton beam of 30$-$50 GeV
rather than to explain approved hadron experiments with kaon
and pion beams.
It is possible to investigate various aspects of hadron and nuclear
structure by using the proton beam.
They include clarification of flavor-dependent antiquark
distributions at large Bjorken $x$ and nucleon spin structure.
The J-PARC facility is expected to play a major role in the studies
of hadron structure and hadronic many-body systems in a different
kinematical region from
RHIC 
and LHC. 

This article consists of the following.
First, the J-PARC facility is introduced. Then, we explain possible
hadron-physics projects with the 30$-$50 GeV proton beam. 
In the last part, our studies on the parton distribution functions
and fragmentation functions are discussed in connection with
possible J-PARC projects. Finally, a summary is given.

\section{J-PARC facility}
\vspace{-0.15cm}

\begin{wrapfigure}{r}{0.48\textwidth}
   \vspace{-0.25cm}
   \begin{center}
       \epsfig{file=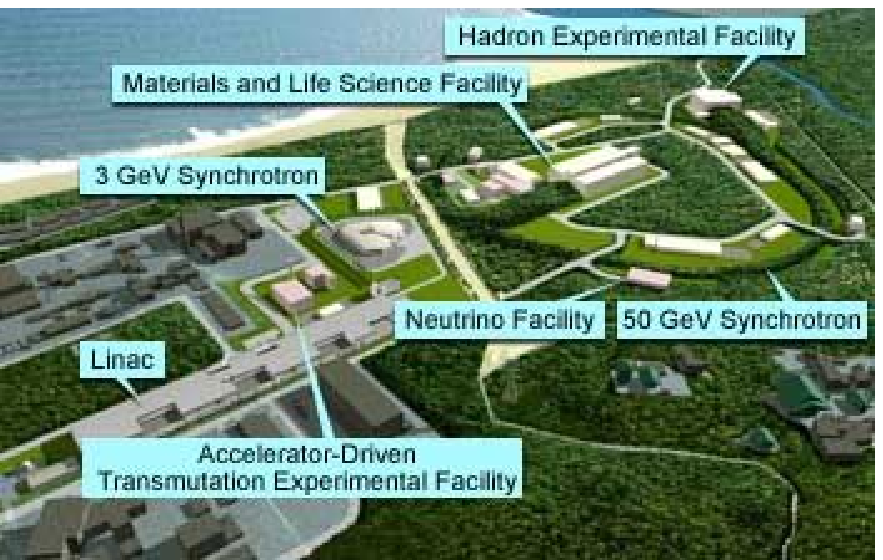,width=0.43\textwidth} \\
   \end{center}
   \vspace{-0.1cm}
       \begin{minipage}[c]{1.5cm}
       \ \ 
       \end{minipage}
       \begin{minipage}[c]{5.0cm}
       \setlength{\baselineskip}{10pt} 
       {\footnotesize Figure 1: Bird's eye view of \\
            the J-PARC facility \cite{j-parc}.}
       \end{minipage}
   \vspace{-0.2cm}
\label{fig:j-parc}
\end{wrapfigure}

The J-PARC accelerator consists of a linac as an injector,
a 3-GeV rapid cycling synchrotron, and a 50-GeV synchrotron
as shown in Fig. 1 \cite{j-parc}.
The J-PARC provides most intense proton beam in the high-energy region
($E>1$ GeV). Other proton accelerators have a beam power less than 0.1 MW,
whereas the J-PARC expects to have 1 MW in the 3-GeV synchrotron
and 0.75 MW in the 50-GeV one.
The J-PARC has three major projects: 
(1) material and life sciences with neutrons and muons
    produced by the 3-GeV proton beam, 
(2) nuclear and particle physics with secondary beams (pions, kaons, 
    neutrinos, and so on) by the 50-GeV proton beam and also with
    protons of the 50-GeV primary beam,
(3) nuclear transmutation by the linac.

\begin{wrapfigure}{r}{0.48\textwidth}
   \vspace{-0.35cm}
   \vspace{-0.25cm}
   \begin{center}
       \epsfig{file=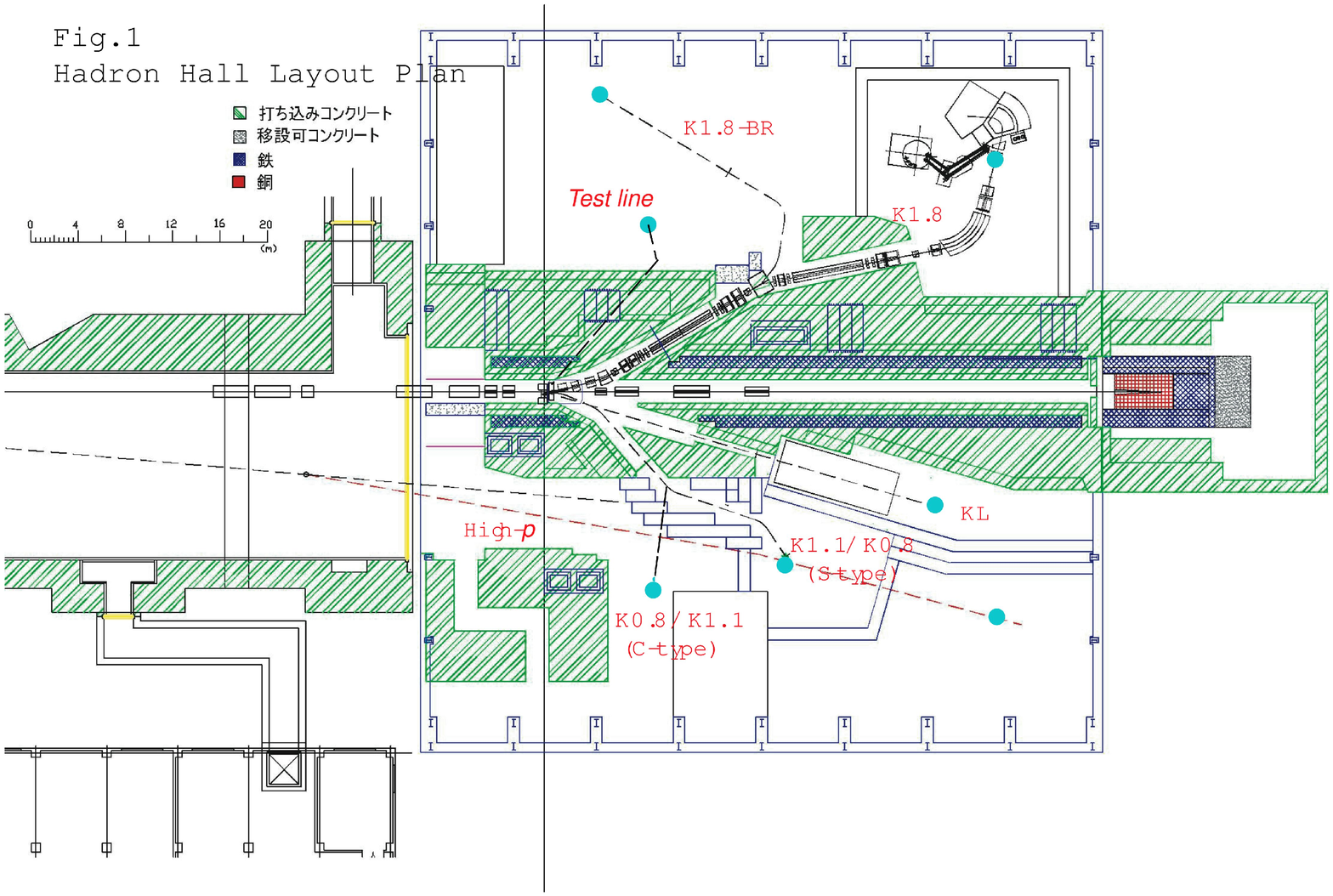,width=0.46\textwidth} \\
   \end{center}
   \vspace{-0.3cm}
       \begin{minipage}[c]{1.0cm}
       \ \ 
       \end{minipage}
       \begin{minipage}[c]{5.5cm}
       \setlength{\baselineskip}{10pt} 
       {\footnotesize Figure 2: Layout of 
                       the hadron hall \cite{j-parc}.}
       \end{minipage}
\label{fig:hadron-hall}
\end{wrapfigure}

Nuclear and hadron-physics experiments will be done
at the hadron experimental facility in Fig. 1 
\cite{j-parc,j-parc-proposals, j-parc-hadron}.
We should mention that there are also hadron topics on neutrino
interactions with the nucleon and nuclei. In this article, we discuss
high-energy hadron projects in the hadron hall, for which
a beam-layout plan is shown in Fig. 2.
The K1.8 is the first beamline which will be
completed. It is intended to have kaons with momentum around
1.8 GeV/$c$ for the studies on strangeness $-2$ hypernuclei
with $\Xi^-$ by $(K^-,K^+)$ reactions.
The K1.1/0.8 beamline is designed for low-momentum stopped
kaon experiments such as the studies of kaonic nuclei.
The neutral kaon beamline (KL) is for studying CP violating
processes such as $K_L \rightarrow \pi^0 \nu\bar\nu$.
The ``High $p$" in Fig. 2 indicates
the high-momentum beamline for 50-GeV protons.
In the beginning stage of J-PARC, the proton beam energy is
30 GeV instead of the designed 50 GeV. At a later stage,
we expect to have 50 GeV energy recovery.

\vspace{-0.15cm}
\section{Hadron physics with primary proton beam}
\vspace{-0.15cm}

There are opportunities to investigate hadron and nuclear
structure at high energies by using the high-momentum beamline
in Fig. 2 \cite{j-parc-hadron,j-parc-conf}.
There are proposals on hadron-mass modifications in a nuclear medium,
Drell-Yan and charm-production processes, 
single spin asymmetries, elastic scattering,
and high-energy spin physics with proton-beam polarization
\cite{j-parc-proposals}. In the following discussions, 
additional topics are explained at the primary proton beamline
from the author's personal point of view without restricting ourselves 
to the proposed experiments.

There are two types of topics: (1) structure functions and related physics
(quark and gluon physics) and (2) hadronic aspects. 

First, the structure functions have both aspects of nonperturbative and
perturbative quantum chromodynamics (QCD). In order to obtain
information about internal structure of hadrons, one needs to subtract
out the perturbative QCD part from cross sections. In order to obtain
reliable results for the nonperturbative part, higher-order QCD corrections
need to be understood.
It is known that such corrections, which are expressed by ``$K$ factors'',
are large in fixed-target experiments. In particular, the center-of-mass
energy is relatively low: $\sqrt{s}=$8 GeV in the beginning and
10 GeV later at the J-PARC. It means that careful estimations
are necessary for the pQCD (perturbative QCD) corrections.
Fortunately, there are significant
developments in the recent years on resummations of soft-gluon radiations
which give rise to large corrections at low energies. It is shown
in Ref. \cite{J-parc-dy-pqcd} that the pQCD corrections converge if
the resummations are properly taken into account at $\sqrt{s}=10$ GeV
in the Drell-Yan processes. 
It means that Drell-Yan measurements are valuable for extracting information
on various parton distribution functions (PDFs) because the perturbative
part is theoretically under control. For charm-production processes, we
still need theoretical studies on such corrections \cite{J-parc-charm-pqcd}.
Conversely, the large corrections mean that J-PARC measurements are
challenging and interesting for pQCD physicists in testing their
high-order calculations.

Second, there are also interesting hadronic topics other than 
the structure functions and related topics.
These projects could be appropriate especially at the 30 GeV energy. 
Possible projects include hadron masses in a nuclear medium, 
transition from hadron degrees of freedom to quark ones 
by elastic $pp$ scattering, color transparency in $(p,2p)$ reactions,
short-range correlation in nuclear force by $(p,2pN)$ reactions, and
generalized parton distributions (GPDs).
These topics are explained in the following.

\vspace{-0.3cm}
\subsection{Partonic structure with primary proton beam}
\vspace{-0.15cm}

\begin{wrapfigure}{r}{0.38\textwidth}
   \vspace{-0.15cm}
   \vspace{-0.25cm}
   \begin{center}
       \epsfig{file=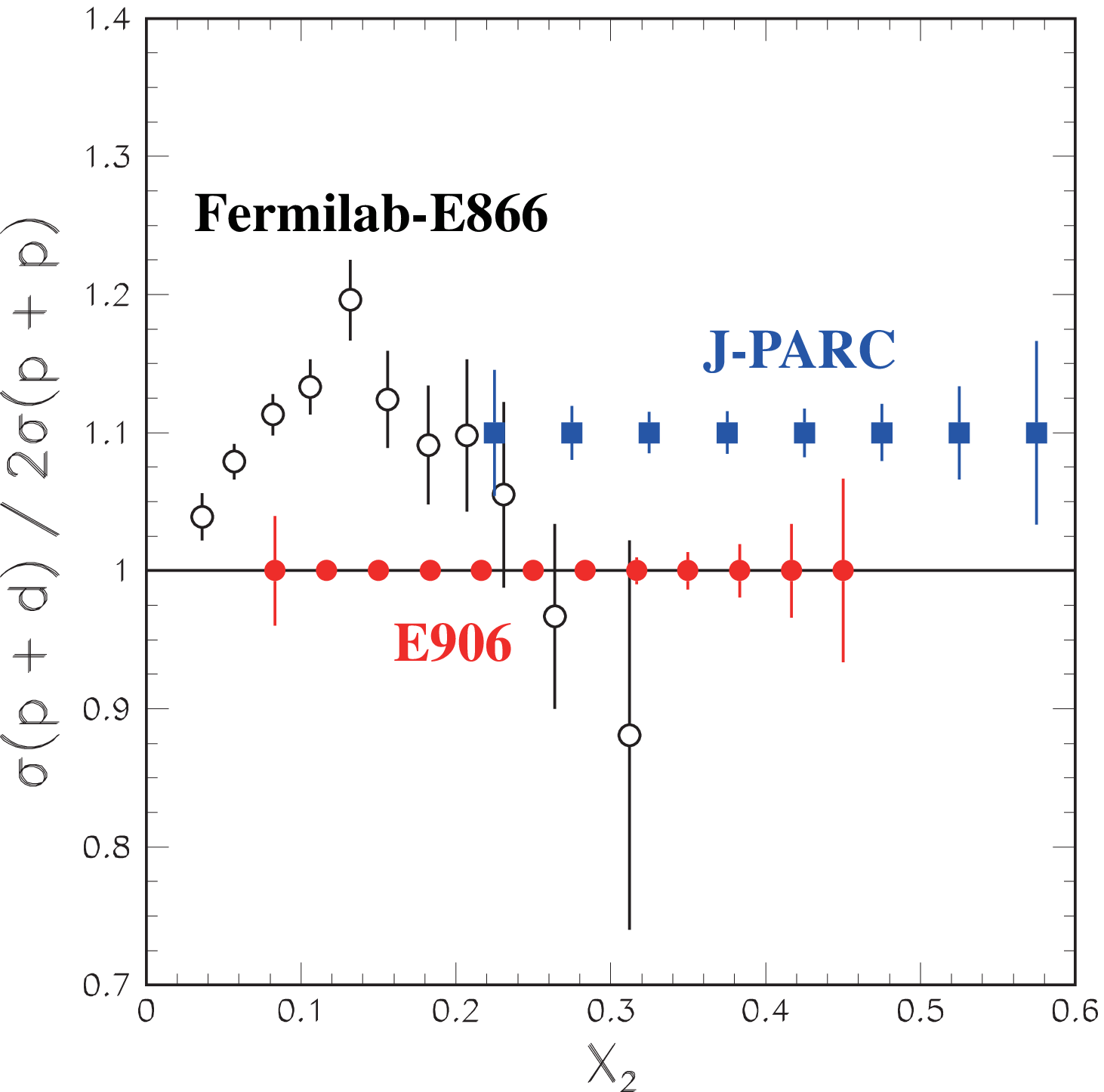,width=0.32\textwidth} \\
   \end{center}
   \vspace{-0.3cm}
       \begin{minipage}[c]{0.5cm}
       \ \ 
       \end{minipage}
       \begin{minipage}[c]{5.0cm}
       \setlength{\baselineskip}{10pt} 
       {\footnotesize Figure 3: Drell-Yan cross section ratio
        $\sigma(p+d)/\sigma(p+p)$ \cite{j-parc-proposals}.}
       \end{minipage}
   \vspace{-0.4cm}
\label{fig:dy}
\end{wrapfigure}

\noindent
\underline{Drell-Yan processes}
\vspace{0.15cm}

Drell-Yan ($pp \rightarrow \mu^+ \mu^- X$) measurements are
important in establishing unpolarized PDFs, especially
antiquark distributions at medium $x$ \cite{dy-peng}.
For example, the Fermilab-E866 measurements in Fig. 3 played
a key role in finding flavor asymmetric antiquark distributions
($\bar u \ne \bar d$). They lead to investigations on 
a nontrivial aspect of nucleon structure \cite{flavor}, 
and physics mechanism could be related to peripheral structure
of the nucleon such as pion clouds. 
The expected measurements by the Fermilab-E906 and J-PARC (50 GeV) 
extend the $x$ region to larger $x$ ($x_{max} \sim 0.6$)
\cite{j-parc-proposals},
where the E866 measurements did not probe. On the other hand,
it is necessary to investigate possible mechanisms theoretically 
to create the difference $\bar u \ne \bar d$ at $x>0.2$.

\vspace{0.20cm}
\noindent
\underline{Charmed-meson productions}
\vspace{0.15cm}

The proton energy is 30 GeV at the initial stage, and
it could be too low to investigate the Drell-Yan process.
One of possible projects with the 30-GeV beam is to study
$J/\psi$ and open-charm production processes 
\cite{j-parc-proposals,J-parc-charm-pqcd}.
The $J/\psi$ production has been discussed as a possible signature
of quark-gluon plasma (QGP) in heavy-ion physics, so that it is 
important to understand its production mechanism. 
Furthermore, the $J/\psi$ and open-charm production
processes probe the gluon distributions in the nucleon and nuclei.
We should note that the gluon distribution in the nucleon
is not determined at large $x$ ($>0.3$),
and it is a major obstacle for finding new physics
by high-$p_T$ jet events in hadron colliders.
The J-PARC is a large-$x$ facility which contributes to
the PDF determination at large $x$ if 
a theoretical description is established for 
the charmed-meson productions.

\vspace{0.20cm}
\noindent
\underline{Parton energy loss}
\vspace{0.15cm}

Recently, parton energy loss became a hot topic in heavy-ion reactions
because it is related to a QGP formation by suppression
of high-$p_T$ mesons. There is also a development on this topic
by the AdS/CFT correspondence. In order to discuss QGP properties,
it is necessary to describe the energy loss properly and to test it
in an independent experiment. 
The Drell-Yan processes with nuclear targets provide a clean
method to investigate the quark energy loss in a cold nuclear medium. 
The energy loss gives rise to nuclear modifications of quark momentum
distributions before a $q\bar q$ annihilation, which results in
changes of Drell-Yan cross sections \cite{j-parc-proposals}.

\begin{wrapfigure}{r}{0.38\textwidth}
   \vspace{-0.15cm}
   \vspace{-0.25cm}
   \begin{center}
       \epsfig{file=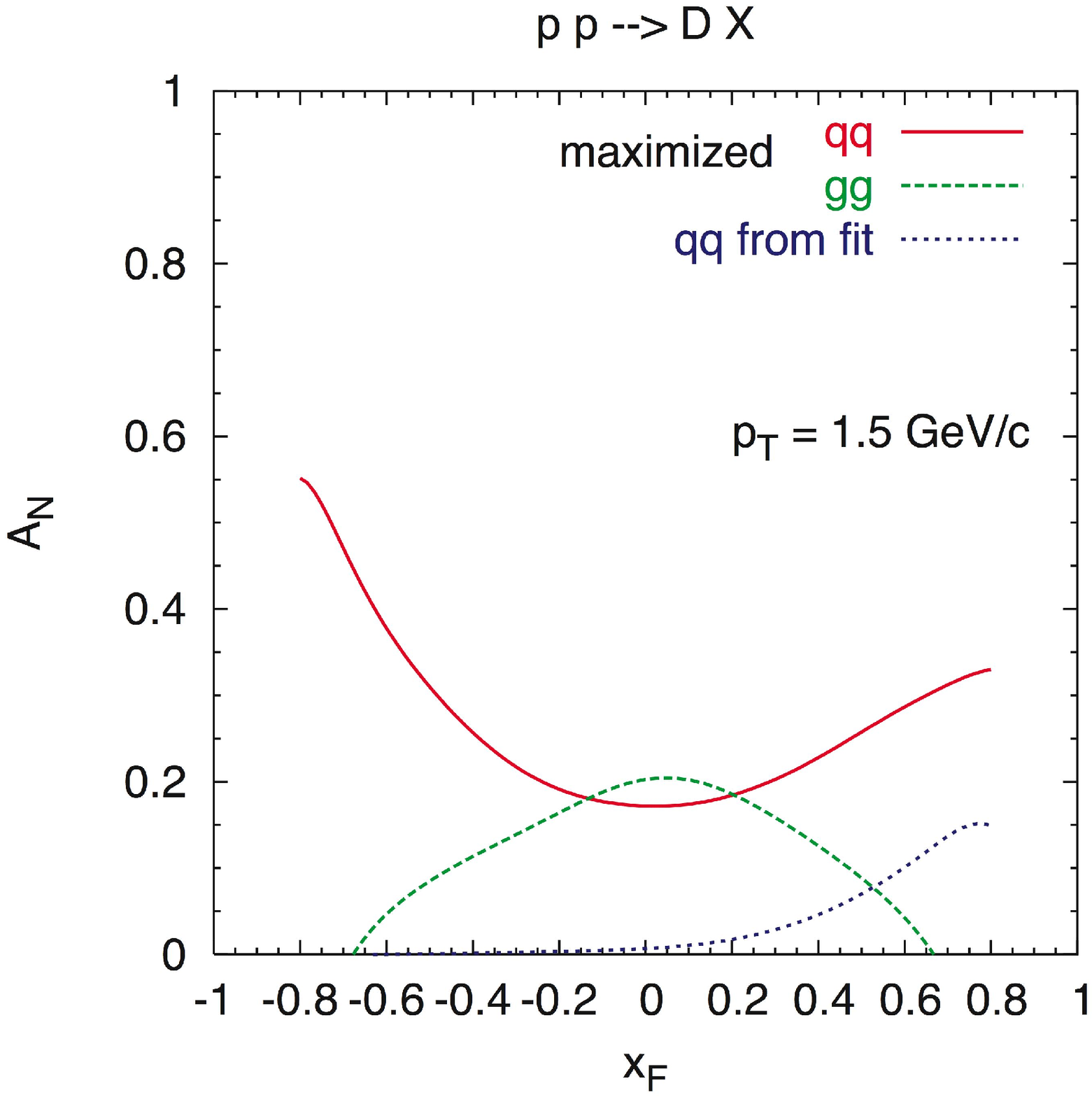,width=0.33\textwidth} \\
   \end{center}
   \vspace{-0.3cm}
       \begin{minipage}[c]{0.5cm}
       \ \ 
       \end{minipage}
       \begin{minipage}[c]{5.0cm}
       \setlength{\baselineskip}{10pt} 
       {\footnotesize Figure 4: Single spin asymmetry for
         D-meson production \cite{j-parc-proposals}.}
       \end{minipage}
   \vspace{-0.3cm}
\label{fig:d-an}
\end{wrapfigure}

\vspace{0.20cm}
\noindent
\underline{Single spin asymmetries}
\vspace{0.15cm}

Single spin asymmetries (SSAs) can be investigated without
proton-beam polarization. 
For example, the SSA for $D$-meson production has been estimated
for the J-PARC in Fig. 4 \cite{j-parc-proposals,single-D}.
Here, the asymmetry could be related to Sivers functions, which
describe unpolarized quarks in the transversely polarized nucleon,
and they are related to angular momenta of quarks.
It is important to note that J-PARC measurements are sensitive to
quark Sivers effects, whereas RHIC ones are to gluon Sivers effects.
The Sivers functions can be also investigated by SSAs
in the $p\vec p$ Drell-Yan process. If a target is transversely polarized,
transversity distributions and Boer-Mulders functions could be also
measured by the SSA in the Drell-Yan \cite{j-parc-proposals}.
The SSA in $pp$ elastic scattering was proposed as a possible experiment 
to investigate more details of observed anomalous asymmetries
in the $p_T \sim$6 GeV region at CERN and BNL by extending
the measurements to $p_T=12$ GeV at J-PARC \cite{j-parc-proposals}.

\vspace{0.20cm}
\noindent
\underline{Tensor structure functions for spin-one hadrons}
\vspace{0.15cm}

There exist new polarized structure functions for spin-1 hadrons
due to their tensor structure nature. There are few experimental
studies on them although spin structure of the spin-1/2 nucleon
has been investigated extensively. At J-PARC without the proton
polarization, tensor polarized distribution functions can be
measured by proton-deuteron Drell-Yan processes with a polarized
deuteron target \cite{pd-dy}. The tensor structure functions are
known as $b_1$ and $b_2$ in the leading twist, and the first
measurement was reported by the HERMES collaboration in 2005
\cite{hermes-b1}. However, experimental errors are large for discussing
$x$ dependence. The $p\vec d$ Drell-Yan measurements at J-PARC 
have an advantage over the HERMES and future lepton-scattering
measurements because antiquark tensor distributions can be measured
for the first time and a sum rule $\int dx b_1(x)=0$
could be studied experimentally \cite{pd-dy,hermes-b1}.
Theoretically, we need to investigate mechanisms to create
the tensor polarization in the quark level.

\vspace{0.20cm}
\noindent
\underline{Fragmentation functions}
\vspace{0.15cm}

Recently, semi-inclusive reactions became important  
for investigating nucleon structure and properties of
quark-hadron matters. In describing hadron-production processes,
fragmentation functions are necessary. A fragmentation function
$D_i^h(z)$ indicates the probability to produce a hadron $h$ 
from a parton $i$ with the energy ratio $z=E_h/E_i$. They have
been determined mainly by hadron-production cross sections
in $e^+e^-$ annihilation. However, the gluon function
has a large uncertainty \cite{ffs-pdfs}, which makes it difficult to 
reliably discuss any hadron-production cross sections in $pp$ and $pA$.
In hadron productions at J-PARC, cross sections are dominated
by gluon-gluon interaction processes, so that they could be
suitable for determining the gluon fragmentation function
at large $z$ ($\sim 1$).

\vspace{-0.3cm}
\subsection{Hadronic structure with primary proton beam}
\vspace{-0.15cm}

We have introduced possible J-PARC projects on partonic structure of hadrons.
At 30 GeV, perturbative corrections are generally large,
so that there are possibilities that
partonic interpretations are difficult for some cross sections.
In the following, a few ideas are given on hadronic projects
which are possible at both 30 and 50 GeV energies
although actual proposals have not been submitted yet
except for the mass modification.

\vspace{0.20cm}
\noindent
\underline{Hadron-mass modifications in nuclear medium}
\vspace{0.15cm}

Chiral symmetry could play an important role in
generating hadron masses because current quark masses
are much smaller than hadron masses.
Chiral symmetry breaking leads to a finite quark condensate,
which is reflected as hadron-mass modifications
in a nuclear medium.
It was proposed to measure the mass modifications of vector mesons
($\rho$, $\omega$, $\phi$) \cite{j-parc-proposals}.
The modifications are theoretically expected 
due to partial restoration of the chiral symmetry.
For example, recent KEK-E325 experimental results indicated 
a 9\% mass shift for $\rho$ and $\omega$. Much accurate data 
will be obtained at J-PARC, so that the chiral dynamics 
will become clear in nuclei. 

\vspace{0.20cm}
\noindent
\underline{Transition from hadron to quark degrees of freedom}
\vspace{0.15cm}

Cross sections for hadron reactions are described by hadron degrees
of freedom (d.o.f.), baryons and mesons, at low energies, whereas
they should be described by quark d.o.f. at high energies.
It is possible to investigate the transition 
from hadron to quark d.o.f. by measuring $pp$ elastic
cross sections at various proton energies.
In the high energy limit, the elastic cross section should be described
by gluon exchanges between constituent quarks, which gives
rise to a counting rule: $d\sigma/dt \sim s^{2-n} f(\theta_{c.m.})$
where $n$ is the total number of interacting elementary particles.
This transition from hadron to quark d.o.f. has been observed in
$\gamma p \rightarrow \pi^+ n$ at $\sqrt{s} \sim$2.5 GeV
\cite{transition}. Similar studies could be done for 
the $pp$ scattering at J-PARC.

\vspace{0.20cm}
\noindent
\underline{Short-range correlation in nuclear force}
\vspace{0.15cm}

Nuclear force has been investigated in terms of
one-boson-exchange processes and also by effective quark models.
Direct relation to QCD is now being studied due to
development of lattice QCD. Short-range repulsion exists
within the distance $r \sim$0.4 fm, and it is especially important
for saturation properties of nuclei.
There is an interesting experimental development on isospin
dependence of the short-range correlation recently. The BNL experiment
$A(p,2pN)X$ and JLab one $A(e,e')X$ at $x>1$ indicated that
$pn$ short-range correlation is twenty times larger than
the $pp$ one \cite{short-range}. The tensor force seems to play
an important role in the short range; however, this unexpected
result needs to be explained theoretically. On the other hand, this result
leads to a modification of neutron-star structure because 
a certain fraction of protons exist in the star. It should be
possible to measure the $A(p,2pN)X$ reaction with better accuracy and
with higher proton momentum (namely shorter range) at J-PARC. 

\vspace{0.20cm}
\noindent
\underline{Color transparency}
\vspace{0.15cm}

At large momentum transfer, a small-size component of the hadron
wave function should dominate in hadron cross sections.
This small-size hadron could pass freely through a nuclear medium,
which is called color transparency. At J-PARC,
the $(p,2p)$ reaction could be investigated for a nuclear target.
Nuclear transparency $T$ is defined by the cross-section ratio
to the nucleonic one: $T=\sigma_A/(A \sigma_N)$. As the hard scale
(energy of the proton) becomes larger, namely as 
the hadron size becomes smaller, the nuclear transparency should
increase. 
Such measurements were done at BNL up to the proton momentum
14 GeV/c \cite{bnl-eva}. The data indicated an interesting
turnover at $p \simeq 9$ GeV/c.
First, we need to establish a theoretical model to explain
the BNL results, and then measurements could be extended
to 30$-$50 GeV/c region at J-PARC for further studies. 

\vspace{0.20cm}
\noindent
\underline{Generalized parton distributions}
\vspace{0.15cm}

Generalized parton distributions (GPDs) contain global information 
on the nucleon structure from low to high-energy region. 
The GPDs become PDFs in the forward limit ($\xi=t=0$),
where $t$ is momentum transfer squared and $\xi$ is a skewdness parameter.
The first moments, namely the GPDs integrated over $x$, are form factors,
and second moments are angular momenta of quarks.
The second moments are especially important because contributions
of angular momenta could solve the nucleon spin issue.
The GPDs have been measured in lepton scattering; however, 
they could be also investigated by proton reactions 
in the ERBL (Efremov-Radyushkin-Brodsky-Lepage) region $-\xi < x < \xi$.
We may investigate $e.g.$ $pp \rightarrow pn\pi$ and 
$pp \rightarrow p\Delta\pi$ for measuring nucleonic and
$N \rightarrow \Delta$ transition GPDs at J-PARC.
A theoretical formalism is now being developed
for such processes by using the 30$-$50 GeV
proton beam \cite{kss08}.

\vspace{-0.3cm}
\subsection{Spin physics with proton beam polarization}
\vspace{-0.15cm}

Feasibility studies indicated that the primary proton beam 
can be polarized at J-PARC \cite{j-parc-proposals}.
By double spin asymmetry experiments, the details of nucleon spin
structure should be investigated. As shown in Fig. 3, the J-PARC
probes the large-$x$ region ($x>0.2$). If the proton-beam polarization
is attained, it is a complementary project to the RHIC-Spin in
investigating a different kinematical region. The RHIC measurements
are generally sensitive to the smaller-$x$ region.

Flavor dependence of the antiquark distributions, $\bar u$ and $\bar d$,
has been investigated by E866 as shown in Fig. 3. 
Flavor dependence of the polarized antiquark distributions is scarcely known
although there are some hints from semi-inclusive DIS experiments for
pion and kaon productions.
J-PARC Drell-Yan measurements on the double-spin asymmetry should be
able to clearly answer the antiquark flavor dependence 
($\Delta\bar u / \Delta\bar d$) at $x>0.2$.
It is important for establishing the origin of nucleon spin
and for testing models of producing the flavor asymmetric 
antiquark distributions. In addition, double spin asymmetry measurements
are possible for hadron ($\pi$, $J/\psi$, ...) and direct-photon productions. 
These measurements should lead to a complete understanding of the nucleon spin 
from small $x$ to large $x$ together with measurements at
other facilities.

\vspace{-0.2cm}
\section{Partonic structure of nucleon and nuclei}
\vspace{-0.15cm}

\begin{wrapfigure}{r}{0.42\textwidth}
   \vspace{-0.25cm}
   \vspace{-0.25cm}
   \begin{center}
       \epsfig{file=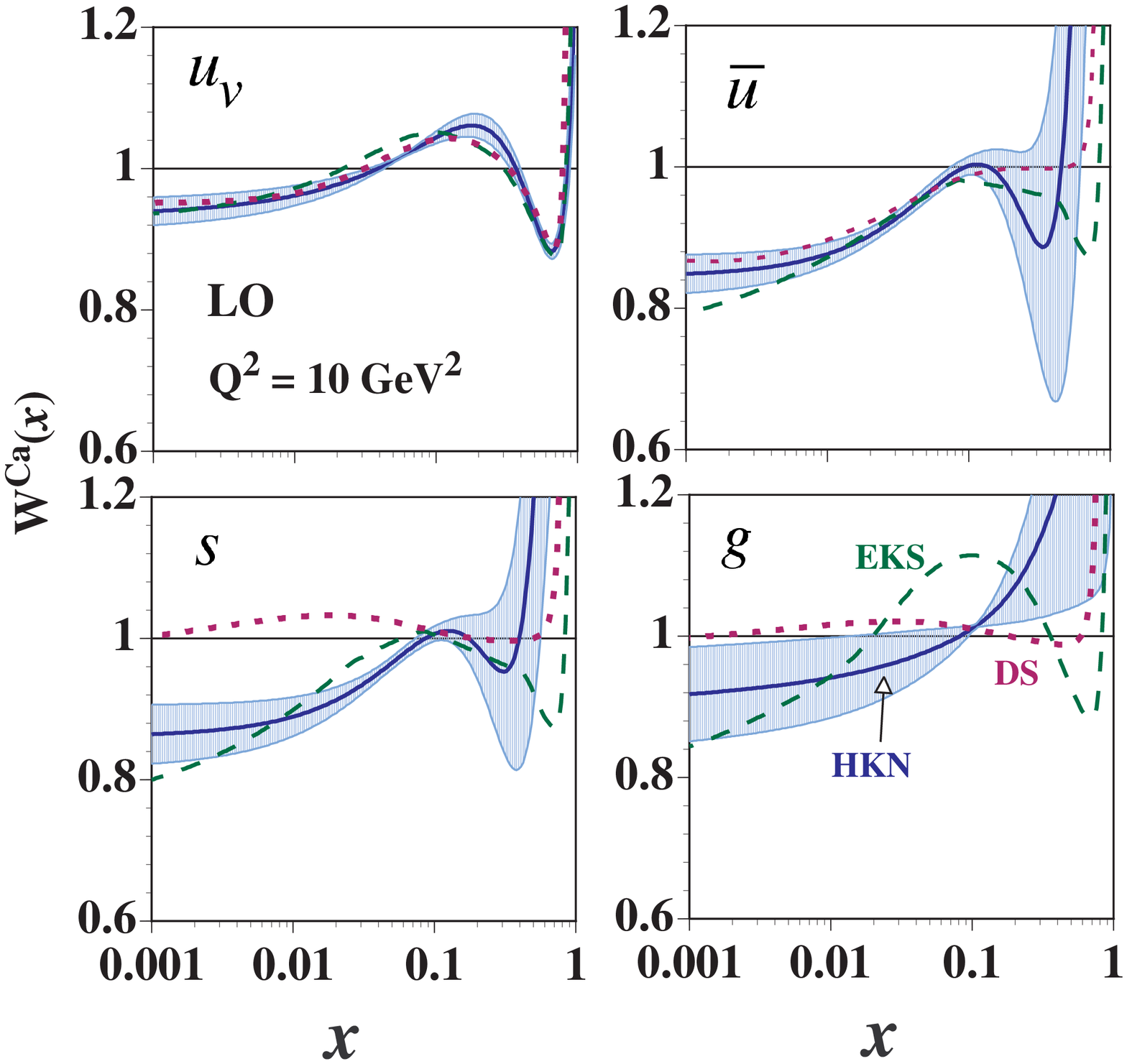,width=0.37\textwidth} \\
   \end{center}
       \vspace{-0.30cm}
       \begin{minipage}[c]{0.6cm}
       \ \ 
       \end{minipage}
       \begin{minipage}[c]{5.3cm}
       \setlength{\baselineskip}{10pt} 
       {\footnotesize Figure 5: Nuclear modifications of PDFs \cite{ffs-pdfs}.}
       \end{minipage}
       \vspace{+0.3cm}
   \begin{center}
       \epsfig{file=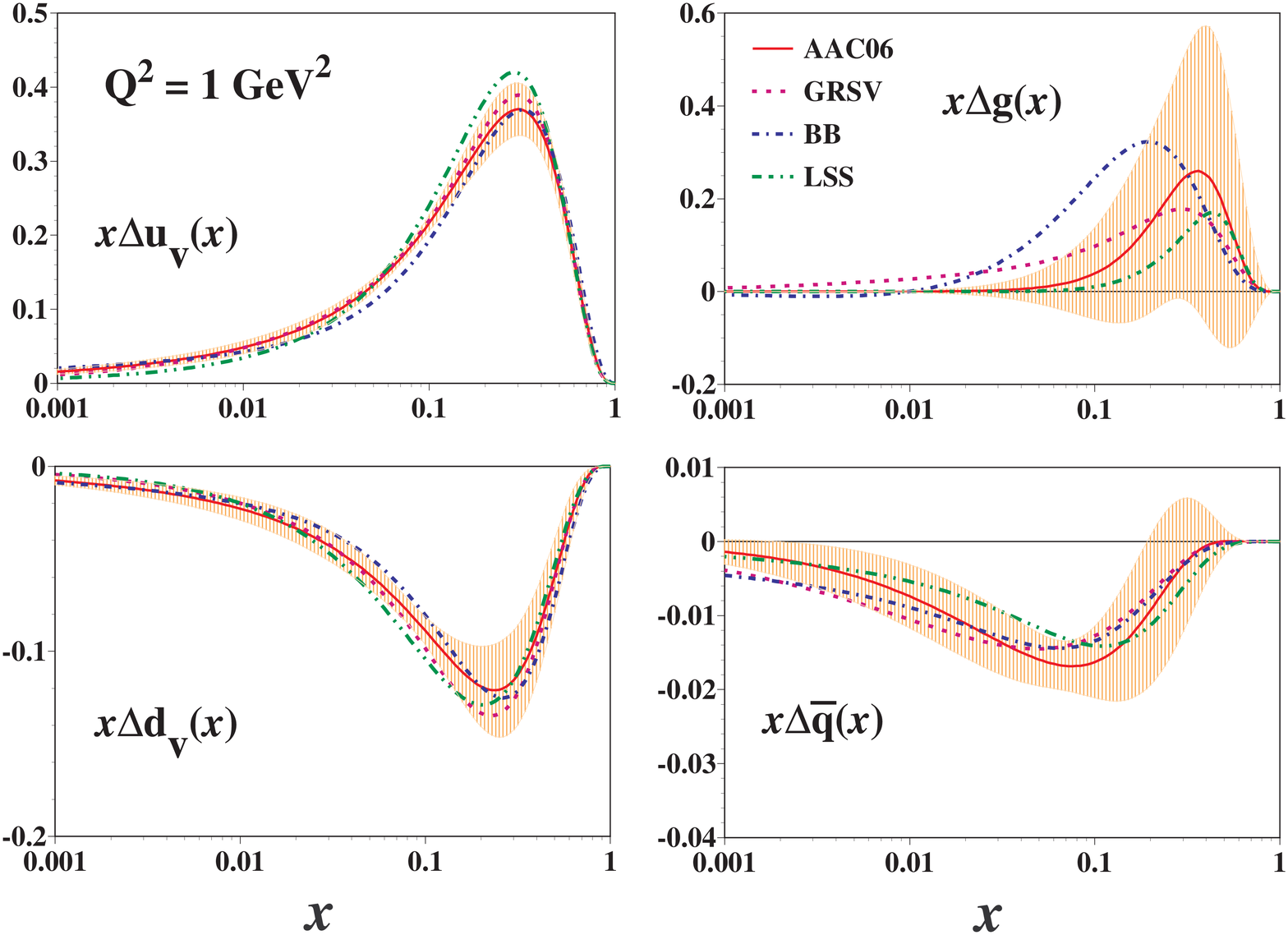,width=0.37\textwidth} \\
   \end{center}
       \vspace{-0.35cm}
       \begin{minipage}[c]{1.0cm}
       \ \ 
       \end{minipage}
       \begin{minipage}[c]{5.0cm}
       \setlength{\baselineskip}{10pt} 
       {\footnotesize Figure 6: Polarized PDFs \cite{ffs-pdfs}.}
       \end{minipage}
       \vspace{+0.0cm}
\end{wrapfigure}

In this last part of this article, our own studies are discussed
on the PDFs in the nucleon and nuclei in connection with
the J-PARC projects.

First, we have been investigating optimum nuclear PDFs by analyzing 
world data on structure function $F_2$ and Drell-Yan cross sections
for nuclei. In Fig. 5, the current situation is shown
for nuclear modifications of the PDFs in the calcium nucleus
at $Q^2$=10 GeV$^2$ \cite{ffs-pdfs}. The uncertainties are shown
by the shaded bands. We notice that antiquark distributions
at $x>0.2$ and gluon distributions are not determined.
By the Drell-Yan and charm-production processes at J-PARC, 
we should be able to determine the antiquark distributions
at $x>0.2$ and possibly also gluon distributions.

Second, the situation of the polarized PDFs is shown in Fig. 6
\cite{ffs-pdfs}. The distributions are not well determined in the polarized
antiquark and gluon distributions. Especially, the gluon distribution
has a large uncertainty band, which is one of the reasons why
the origin of the nucleon spin is not clarified.
If the proton-beam polarization is attained at J-PARC in future, 
these distributions can be investigated by the double spin asymmetries
in Drell-Yan and hadron-production processes.

Third, fragmentation functions (FFs) could be investigated by
hadron-production processes at J-PARC. The FFs have been determined
mainly by hadron productions in $e^+e^-$ annihilation.
We found in Ref. \cite{ffs-pdfs} that gluon and light-quark FFs 
have large uncertainties.
The gluon function should be determined by hadron facilities
such as RHIC, LHC, and J-PARC because gluon subprocesses dominate
the small-$p_T$ part of hadron-production cross sections.
In particular, the lower-energy J-PARC probes the very large-$z$
region ($z \sim 1$), so that measurements should be important
for establishing fragmentation processes from a gluon.

\vspace{-0.3cm}
\section{Summary}
\vspace{-0.15cm}

The hadron-physics experiments will start at J-PARC in 2009, and
they are important for new developments in hadron and nuclear physics.
The first project is on new many-body systems with strangeness
by using pion and kaon beams.
In this article, we focused on other hadron-physics projects  
at high energies by using the 30$-$50 GeV primary proton beam.

The proposed Drell-Yan and charm-production experiments are
valuable for clarifying the flavor dependence of antiquark
distributions in the nucleon at $x>0.2$, nuclear antiquark
distributions, and possibly also the gluon distributions 
at large $x$ in the nucleon and nuclei. The parton-energy loss
can be studied in the Drell-Yan.
From single spin asymmetry measurements, it could be possible
to learn about orbital angular momentum contributions to the nucleon spin. 
It is important to measure mass modifications of vector mesons 
in a nuclear medium for investigating chiral dynamics.
There are other interesting projects in addition to these proposed experiments.
They include new spin-1 structure, fragmentation functions, transition from
hadron to quark d.o.f., short-range correlation in nuclear force,
color transparency, and GPDs.
If the proton beam is polarized, nucleon spin structure can be
studied in details, for example, on the flavor dependence of
polarized antiquark distributions.
In the last part, we explained the current status of the PDFs and
fragmentation functions, which are related to the J-PARC projects,
by global analyses of experimental data.
The J-PARC will be one of leading facilities in hadron
and nuclear physics. 




\begin{thebibliography}{9}
\vspace{-0.12cm}
\bibitem{j-parc} http://j-parc.jp/index-e.html.
                S. Sawada, Nucl. Phys. {\bf A782}, 434 (2007).
\bibitem{j-parc-proposals} http://www.j-parc.jp/NuclPart/Proposal\_e.html.
                High-energy hadron projects are in
                  the proposals: P04, J. Chiba {\it et al.}   (2006); 
                                 P23, A. W. Chao {\it et al.} (2007); 
                                 P24, M. Bai {\it et al.}     (2007).
               The proposal on the mass modification is 
                                 P16, S. Yokkaichi {\it et al.} (2006).
\bibitem{j-parc-hadron} S. Kumano, Nucl. Phys. {\bf A782}, 442 (2007).
  For overview talks on hadron physics at J-PARC, see 
  http://www.pg.infn.it/hadronic06/; 
  http://inwpent5.ugent.be/workshop07/;
  http://j-parc.jp/NP08/.
\bibitem{j-parc-conf}
  For possible topics on hadron physics at J-PARC, see slides at
  http://www-conf.kek.jp/J-PARC-HS05 \\ /program.html;
  http://www-conf.kek.jp/NP\_JPARC/program.html;
  http://j-parc.jp/NP08/.
\bibitem{J-parc-dy-pqcd} H. Yokoya and W. Vogelsang, hep-ph/0607043,
       pp. 723-736 in Proceedings of 14th International Workshop 
       on Deep Inelastic Scattering, World Scientific (2007).
\bibitem{J-parc-charm-pqcd}
  J.-W. Qiu, talk at this workshop,
  http://cdsagenda5.ictp.trieste.it/full\_display.php?email=0\&ida=a07151;
  M. Stratmann in http://www-conf.kek.jp/hadron08/hehp-jparc/.
\bibitem{dy-peng} J.-C. Peng, talk at this workshop, 
                              arXiv:0807.3538 [nucl-ex].
\bibitem{flavor}  S. Kumano, Phys. Rept. {\bf 303}, 183 (1998);
                  G. T. Garvey and J.-C. Peng,
                       Prog. Part. Nucl. Phys. {\bf 47}, 203 (2001).                
\bibitem{single-D} M. Anselmino {\it et al.},
                       Phys. Rev. D {\bf 70}, 074025 (2004).
\bibitem{pd-dy} S. Hino and S. Kumano, 
                       Phys. Rev. D {\bf 59}, 094026 (1999);
                                  D {\bf 60}, 054018 (1999);
                F. E. Close and S. Kumano, 
                       Phys. Rev. D {\bf 42}, 2377 (1990).
\bibitem{hermes-b1} A. Airapetian {\it et al.}, 
                       Phys. Rev. Lett. {\bf 95}, 242001 (2005).
\bibitem{ffs-pdfs} M. Hirai {\it et al.}, 
                       Phys. Rev. D {\bf 75}, 094009 (2007); 
                                  C {\bf 76}, 065207 (2007);
                                  D {\bf 74}, 014015 (2006).
\bibitem{transition} L. Y. Zhu {\it et al.}, 
                       Phys. Rev. Lett. {\bf 91}, 022003 (2003).
\bibitem{short-range} E. Piasetzky {\it et al.},
                         Phys. Rev. Lett. {\bf 97}, 162504 (2006);
                      R. Shneor {\it et al.},
                         Phys. Rev. Lett. {\bf 99}, 072501 (2007). 
\bibitem{bnl-eva} J. Aclander {\it et al.}, 
                       Phys. Rev. C {\bf 70}, 015208 (2004). 
\bibitem{kss08} S. Kumano, M. Strikman, and K. Sudoh, research in progress.
\end{thebibliography}
\end{document}